\documentclass[seceq]{ptptex}

\usepackage{graphicx}

\title{
The Derivation of  the Exact Internal Energies for Spin Glass Models by Applying the Gauge Theory to the Fortuin-Kasteleyn Representation%
}

\author{
Chiaki \textsc{Yamaguchi}%
}

\inst{
Kosugichou 1-359, Kawasaki 211-0063, Japan
}

\abst{
 We derive the exact internal energies and the rigorous upper bounds of specific
 heats for several spin glass models
 by applying the gauge theory to the Fortuin-Kasteleyn representation
 which is a representation based on a percolation picture for spin-spin correlation.
 The results are derived on the Nishimori lines which are special lines on the phase diagrams.
 As the spin glass models, the $\pm J$ Ising model and 
 a Potts gauge glass model are studied.
 The present solutions agree with the previous solutions.
 The derivation of the solutions by 
 the present method must be useful for understanding the relationship
 between the percolation picture for spin-spin correlation
 and the physical quantities on the Nishimori line.
}

\begin{document}
\maketitle

\section{Introduction}

The theoretical studies of spin glasses have been 
 widely done \cite{KR}.
 There are special lines
 on the phase diagrams for several spin glass models,
 where the lines are called the Nishimori line \cite{N, MH, NS}.
 It is known that several physical quantities and several bounds
 for physical quantities
 are exactly calculated on the Nishimori line 
 by using gauge transformations.
 The exact internal energies and the rigorous upper bounds of 
 specific heats for several spin glass models 
 have already been derived on the Nishimori line \cite{N, MH, NS}.
 The aim of this article is to derive the exact internal energies and
 the rigorous upper bounds of specific heats for several spin glass models
 by applying the gauge theory to the Fortuin-Kasteleyn representation.
 As the spin glass models, the $\pm J$ Ising model \cite{N, MH} and 
 a Potts gauge glass model \cite{NS} are studied.
 The Potts  gauge glass model is a more complex version of the  $\pm J$ Ising model.

 The Fortuin-Kasteleyn
 representation is a representation based on a percolation picture for
 spin-spin correlation \cite{KF, FK, KG}.
 By using the Fortuin-Kasteleyn representation,
 spin-spin correlations are directly treated.
 In the previous methods,
 the solutions have been directly calculated from the Boltzmann factor \cite{N, MH, NS}.
 Instead, in the present method,
 the solutions are directly calculated from 
 the Fortuin-Kasteleyn representation.

In this article,
 gauge transformations are used.
 The gauge transformations are treated in Refs.~\citen{N, MH, NS, T, Y, Y2} for example.
 It is known that 
 a gauge transformation has no effect on thermodynamic quantities \cite{T}.

This article is organized as follows.
 The Fortuin-Kasteleyn representation is briefly explained in \S\ref{sec:2}.
 The solutions for the $\pm J$ Ising model are obtained in \S\ref{sec:3}.
 The solutions for a Potts gauge glass model are obtained in \S\ref{sec:4}.
 This article is summarized in \S\ref{sec:5}.
 Appendices A and B are attached in order to make this article
 self-contained.

\section{The Fortuin-Kasteleyn representation} \label{sec:2}

We briefly explain the Fortuin-Kasteleyn representation \cite{KF, FK}.
 The Fortuin-Kasteleyn representation introduces auxiliary variables called graph G.
 The graph $G$ is a state by the weights between spins which are directly connected by the interaction.
 The partition function $Z$ is expressed in the double
 summation over state $S$ and graph $G$ as \cite{KG}
\begin{equation}
 Z = \sum_{S, G} V (G) \Delta (S, G) \, ,
\end{equation}
 where 
 $\Delta (S, G)$ is a function that takes
 the value one when $S$ is compatible to $G$
 and takes the value zero otherwise.
 A bond that probabilistically connects two spins by the weight of graph is especially called the active bond.
 A graph consists of a set of active bonds.
The active bond is fictitious, and is used in order to generate a cluster composed of spins, which is often referred to as the Fortuin-Kasteleyn cluster.
 $V (G)$ is the weight for the graph $G$.
 The partition function $Z$ is expressed as $\sum_G W(G)$,
 where $W(G) = V (G) \sum_S \Delta (S, G) $.
 This partition function is expressed in the summation over graph $G$
 instead of state $S$. 
 Since the weight for the graph $G$ is used,
 the study of the percolation problem of the graph $G$
 is expected to have a physical significance.
 This representation for graph $G$ is called the Fortuin-Kasteleyn representation.

 We define the number of active bonds as $n_b$.
 The number of states for the active bond number, $\Omega (n_b )$,
 is given by 
\begin{equation}
 \Omega (n_b ) = \sum_{\{ G | n_b (G) = n_b \}}
 \sum_S \Delta (S, G) \, .
\end{equation}
 By using the $\Omega (n_b )$,
 the partition function is expressed as
\begin{equation}
 Z = \sum_{n_b = 0}^{N_B} \Omega (n_b ) V (n_b ) \, , \label{eq:Z} 
\end{equation}
 where $V (n_b )$ is the weight for the active bond number,
 and $N_B$ is the number of nearest-neighbor pairs in the whole system.

 If two spins are on the same cluster, the two spins are correlated.
 If two spins are not on the same cluster, the two spins are not correlated.
 In the ferromagnetic Ising model,
 the percolation transition point of the Fortuin-Kasteleyn cluster
 agrees with the phase transition point \cite{CK}.
 On the other hand, in the $\pm J$ Ising model,
 the percolation transition point of the Fortuin-Kasteleyn cluster
 disagrees with the phase transition point \cite{MNS}.
 Instead, it is pointed out that, in the $\pm J$ Ising model,
 there is a possibility that
 the percolation transition point of the Fortuin-Kasteleyn cluster
 agrees with a dynamical transition point \cite{ACP}.
 For the applications of the Fortuin-Kasteleyn representation,
 the Swendsen-Wang algorithm \cite{SW} is probably the prime example.
 This algorithm is a Markov chain Monte Carlo method.
 By performing this algorithm,
 the Fortuin-Kasteleyn clusters are generated,
 and the states on each cluster are simultaneously updated.
 This algorithm produces a faster thermal equilibration when
 this algorithm is applied to the ferromagnetic Ising model \cite{SW}.
 In this article, we concentrate ourselves on the number of the active bonds,
 which generate the Fortuin-Kasteleyn clusters,
 and the fluctuation of the number of active bonds.

\section{The $\pm J$ Ising model and the present results} \label{sec:3}

The Hamiltonian for the $\pm J$ Ising model, ${\cal H}$, 
 is given by  \cite{KR, N, MH}
\begin{equation}
 {\cal H} = - \sum_{\langle i, j \rangle} J_{i, j} S_i S_j \, ,
\end{equation}
 where $\langle i, j \rangle$ denotes nearest-neighbor pairs, $S_i$ is
 a state of the spin at the site $i$, and $S_i = \pm 1$.
 $J_{i, j}$ is a strength of the exchange interaction between the spins at the sites $i$ and $j$.
 The value of $J_{i, j}$ is given with a distribution $P (J_{i, j})$.
 The distribution $P (J_{i, j})$ is given by
\begin{equation}
 P (J_{i, j})
 = p \, \delta_{J_{i, j}, J} + (1 - p) \, \delta_{J_{i, j}, - J} \, ,
 \label{eq:PpmJJij}
\end{equation}
 where $J > 0$, and $\delta$ is the Kronecker delta.
 $p$ is the probability that the interaction
 is ferromagnetic, and $1 - p$ is
 the probability that the interaction is antiferromagnetic. 
 By using Eq.~(\ref{eq:PpmJJij}), the distribution $P (J_{i, j})$ 
 is written as \cite{N, MH, Y}
\begin{equation}
 P (J_{i, j}) = \frac{e^{\beta_{\rm P} J_{i, j}}}
{2 \cosh (\beta_P J)} \, , \quad J_{i, j} = \pm J \label{eq:PJij} \, ,
\end{equation}
 where $\beta_P$ is given by \cite{N, MH, Y}
\begin{equation}
 \beta_P = \frac{1}{2 J} \ln \frac{p}{1-p} \, . \label{eq:betaPpmJ}
\end{equation}
 When the value of $\beta_P$ is consistent with the value of 
 the inverse temperature $\beta$, 
 the line on the phase diagram  for the temperature $T$ and $p$,
 where Eq.~(\ref{eq:betaPpmJ}) is satisfied,
 is called the Nishimori line.

A gauge transformation \cite{N, MH, T, Y} given by
\begin{equation}
 J_{i, j} \to J_{i, j} \sigma_i \sigma_j \, , \quad S_i \to S_i \sigma_i 
 \label{eq:GaugeT} 
\end{equation}
 is used where $\sigma_i = \pm 1$.
 By using the gauge transformation,
 the Hamiltonian  ${\cal H}$ part becomes ${\cal H} \to {\cal H}$, 
 and the distribution $P (J_{i, j})$ part becomes
\begin{eqnarray}
 \prod_{\langle i, j \rangle} P (J_{i, j})
 &=& \frac{e^{\beta_{\rm P} \sum_{\langle i, j \rangle} J_{i, j}}}
{[2 \cosh (\beta_P J)]^{N_B}}  \nonumber \\
 &\to& \frac{\sum_{\{ \sigma_i \}} e^{\beta_{\rm P}
 \sum_{\langle i, j \rangle}
 J_{i, j} \sigma_i \sigma_j}}
{2^ N [2 \cosh (\beta_P J)]^{N_B}} \, ,
 \label{eq:PJij2}
\end{eqnarray}
 where $N$ is the number of sites.

 For the $\pm J$ Ising model,
 $V (n_b ) $ is given by
\begin{equation}
 V (n_b ) = (e^{2 \beta J} - 1)^{n_b} e^{- N_B \beta J} \, ,
 \label{eq:VpmJ}
\end{equation}
 where $\beta = 1 / k_B T$, $T$ is the temperature,
 and $k_B$ is the Boltzmann constant.
 The way of deriving Eq.~(\ref{eq:VpmJ}) is described in Appendix A.
We define the probability for putting the active bond
 as $P_{\rm FK}$. The value of $P_{\rm FK}$ depends on
 the exchange interaction and the states of spins \cite{KF, FK, KG, ACP, Y, Y2}. 
 For the $\pm J$ Ising model, $P_{\rm FK}$ is given by \cite{ACP, Y}
\begin{equation}
 P_{\rm FK} (S_i, S_j, J_{i, j}) = 1 -
 e^{- \beta J_{i, j} S_i S_j - \beta |J_{i, j}|} \, . \label{eq:PFKpmJ}
\end{equation}
 The way of deriving Eq.~(\ref{eq:PFKpmJ}) is also described in Appendix A.
 By using the gauge transformation, the $P_{\rm FK}$ part becomes $P_{\rm FK} \to P_{\rm FK}$.

The internal energy $E$ is given by
\begin{equation}
 E = - \frac{\partial}{\partial \beta}
 [ \ln Z ]_R \, , \label{eq:energy}
\end{equation}
 where $[ \, ]_R$ denotes the random configuration average.
 By using Eqs.~(\ref{eq:Z}), (\ref{eq:VpmJ}) and (\ref{eq:energy}), we obtain
\begin{equation}
 E  = N_B  J - \frac{2 J e^{2 \beta J}}{e^{2 \beta J} - 1}
 [ \langle n_b \rangle_T ]_R \, , \label{eq:energy_pmJ} 
\end{equation}
 where $\langle \, \rangle_T$ denotes the thermal average.
 $[\langle n_b \rangle_T ]_R$ is given by
\begin{equation}
 [ \langle n_b \rangle_T ]_R = \sum_{\langle i, j \rangle} 
 [\langle P_{\rm FK} (S_i, S_j, J_{i, j}) \rangle_T ]_R  \, . \label{eq:nbTRpmJ1}
\end{equation}
 When $\beta = \beta_P$,
 $[\langle P_{\rm FK} (S_i, S_j, J_{i, j}) \rangle_T ]_R$ is obtained
 by using the gauge transformation as \cite{Y}
\begin{eqnarray}
  & & [\langle P_{\rm FK} (S_i, S_j, J_{i, j}) \rangle_T ]_R
  \nonumber \\
  &=& \sum_{ \{ J_{l, m} \}}
 \prod_{\langle l, m \rangle} P (J_{l, m}) 
  \frac{\sum_{\{ S_l \} } 
 P_{\rm FK} (S_i, S_j, J_{i, j})
 \, e^{- \beta_P {\cal H} (\{ S_l \}, \{ J_{l, m} \}) }}
 {\sum_{\{ S_l \} } 
 e^{- \beta_P {\cal H} (\{ S_l \}, \{ J_{l, m} \})}} \nonumber \\
  &=& \frac{1}{2^N [2 \cosh (\beta_P J)]^{N_B}} \sum_{ \{ J_{l, m} \}}
 \sum_{\{ S_l \} }
 P_{\rm FK} (S_i, S_j, J_{i, j})
 \, e^{- \beta_P {\cal H} (\{ S_l \}, \{ J_{l, m} \})} \nonumber \\ 
 &=& \tanh ( \beta_P J) \, , 
 \label{eq:PFKTRpmJ}
\end{eqnarray}
 where $\beta_P = 1 / k_B T_P$, $T_P$ is the temperature on the Nishimori line.
 By using Eqs.~(\ref{eq:nbTRpmJ1}) and (\ref{eq:PFKTRpmJ}),
 we obtain
\begin{equation}
 [ \langle n_b \rangle_T ]_R = N_B \tanh (\beta_P J) \, . \label{eq:nbTRpmJ2}
\end{equation}
 By using Eqs.~(\ref{eq:energy_pmJ}) and (\ref{eq:nbTRpmJ2}),
 the internal energy $E$ is obtained as
\begin{equation}
 E = - N_B  J \tanh (\beta_P J) \, .
\end{equation}
 This solution is exact, and is equivalent to
 the solution in Ref.~\citen{N}.

The specific heat $C$ is given by
\begin{equation}
 C = k_B \beta^2 \frac{\partial^2}{\partial \beta^2}
 [ \ln Z ]_R \, . \label{eq:C}
\end{equation}
By using Eqs.~(\ref{eq:Z}), (\ref{eq:VpmJ}) and (\ref{eq:C}),
 we obtain
\begin{eqnarray}
 C &=& k_B (\beta J)^2 {\rm cosech}^2 (\beta J)
 \{ - [\langle n_b \rangle_T ]_R \nonumber \\
 &+& e^{2 \beta J} ([\langle n_b^2 \rangle_T ]_R
 - [ \langle n_b \rangle_T^2 ]_R ) \} \, .
 \label{eq:CpmJ}
\end{eqnarray}
 $[\langle n_b^2 \rangle_T ]_R$ is given by
\begin{eqnarray}
 & & [\langle n_b^2 \rangle_T ]_R \nonumber \\
 &=& \sum_{\langle i, j \rangle} \sum_{\langle k, l \rangle}
 [\langle P_{\rm FK} (S_i, S_j, J_{i, j})
 P_{\rm FK} (S_k, S_l, J_{k, l})
 (1 - \delta_{i, k} \delta_{j, l} - \delta_{i, l} \delta_{j, k}  )
 \nonumber \\
 &+& P_{\rm FK} (S_i, S_j, J_{i, j})
 (\delta_{i, k} \delta_{j, l} + \delta_{i, l} \delta_{j, k} ) \rangle_T ]_R \, .
\end{eqnarray}
 By performing a similar calculation with 
 the calculation in Eq.~(\ref{eq:PFKTRpmJ}), we obtain
\begin{eqnarray}
 [\langle n_b^2 \rangle_T ]_R &=& N_B (N_B - 1) \tanh^2 (\beta_P J) \nonumber \\
 &+& N_B \tanh (\beta_P J) \, . \label{eq:nb2TR}
\end{eqnarray}
 By applying the Cauchy-Schwarz inequality,
 we obtain
\begin{equation}
 [ \langle n_b \rangle^2 ]_R \ge [ \langle n_b \rangle ]^2_R
 = N_B^2 \tanh^2 (\beta_P J) \, .
 \label{eq:nb2TR2}
\end{equation}
 Therefore, by using Eqs.~(\ref{eq:nbTRpmJ2}),
 (\ref{eq:CpmJ}), (\ref{eq:nb2TR}) and (\ref{eq:nb2TR2}),
 we obtain the upper bound of the specific heat $C$ as
\begin{equation}
 C \le k_B N_B (\beta_P J )^2 {\rm sech}^2 (\beta_P J) \, .
\end{equation}
 This solution is rigorous, and is equivalent to
 the solution in Refs.~\citen{N, MH}.

\section{A Potts gauge glass model and the present results} \label{sec:4}

The Hamiltonian for a Potts gauge glass model, ${\cal H}$, 
 is given by \cite{NS}
\begin{equation}
 {\cal H} = - \frac{J}{q} \sum_{\langle i, j \rangle}
 \sum_{r_{i, j} = 1}^{q - 1} e^{\frac{2 \pi i}{q} ( \nu_{i, j} + q_i - q_j) r_{i, j}} \, ,
 \label{eq:HamiltonianPGG}
\end{equation}
 where $q_i$ is
 a state of the spin at the site $i$, and $q_i = 0, 1, \ldots, q - 1$.
 $\nu_{i, j}$ is a variable related to the strength
 of the exchange interaction between the spins at the sites $i$ and $j$,
 and $\nu_{i, j} = 0, 1, \ldots, q - 1$.
 $q$ is the total number of states that a spin takes.
 The value of $\nu_{i, j}$ is given with a distribution $P (\nu_{i, j} )$.
 The distribution $P ( \nu_{i,j} )$
 is given by
\begin{equation}
 P ( \nu_{i, j} ) = p \, \delta_{ \nu_{i, j}, 0} + \frac{1 - p}{q - 1}
 (1 - \delta_{ \nu_{i, j}, 0} ) \, . \label{eq:Pnuij}
\end{equation}
 The normalization of $P ( \nu_{i, j} )$ is given by
\begin{equation}
 \sum^{q - 1}_{\nu_{i, j} = 0}
 P ( \nu_{i, j} ) = 1 \, . \label{eq:pnuijnorm}
\end{equation}
 When $\nu_{i, j} = 0$ for all $(i, j)$ pairs,
  the model becomes the ferromagnetic Potts model.
 When $q = 2$, the model becomes the $\pm J$ Ising model.
 By using Eqs.~(\ref{eq:Pnuij}) and (\ref{eq:pnuijnorm}),
 the distribution $P ( \nu_{i, j} )$ is written as \cite{NS, Y2}
\begin{equation}
 P ( \nu_{i, j} ) =  A e^{\frac{\beta_{\rm P} }{q}
 \sum_{r_{i, j} = 1}^{q - 1} J^{(r_{i, j})}_{i, j} ( \nu_{i, j} )} \, ,
 \label{eq:Pnuij2}
\end{equation}
 where $A$ and $\beta_P$ are given by \cite{NS, Y2}
\begin{eqnarray}
 A &=& \frac{1}{e^{\frac{\beta_P J}{q} (q - 1)}
 + (q - 1) e^{- \frac{\beta_P J}{q}}}
  \, , \label{eq:A} \\
 \beta_P &=& \frac{1}{J}
 \ln \biggl[ p \biggl( \frac{q - 1}{1-p} \biggr)
 \biggr] \label{eq:betaPnuij}
\end{eqnarray}
 respectively.
 When the value of $\beta_P$ is consistent with the value of 
 the inverse temperature $\beta$, 
 the line on the phase diagram  for the temperature $T$ and $p$,
 where Eq.~(\ref{eq:betaPnuij}) is satisfied,
 is called the Nishimori line.

We use representations: $\lambda_i = e^{\frac{2 \pi i}{q} q_i}$ and 
 $J^{(r_{i, j})}_{i, j} = J e^{\frac{2 \pi i}{q} \nu_{i, j} r_{i, j}}$.
 A gauge transformation \cite{NS, Y2} given by
\begin{equation}
 J^{( r_{i, j} ) }_{i, j} \to J^{( r_{i, j} )}_{i, j} \mu^{q - r_{i, j}}_i
 \mu^{r_{i, j}}_j \, , \quad \lambda_i \to \lambda_i \mu_i  \label{eq:GaugeT2} 
\end{equation}
 is used where $\mu_i = e^{\frac{2 \pi i}{q} \tilde{q}_i}$,
 $\tilde{q}_i$ is an arbitrary value for the spin state at the site $i$,
 and $\tilde{{q}_i} = 0, 1, \ldots, q - 1$.
 By using the gauge transformation,
 the Hamiltonian  ${\cal H}$ part becomes ${\cal H} \to {\cal H}$,
 and the distribution $P (\nu_{i, j})$ part becomes
\begin{eqnarray}
  \prod_{\langle i, j \rangle} P (\nu_{i, j})
  &=& 
 A^{N_B} e^{\frac{\beta_{\rm P} }{q} \sum_{\langle i, j \rangle}
 \sum_{r_{i, j} = 1}^{q - 1} J^{(r_{i, j})}_{i, j} ( \nu_{i, j} )} 
 \nonumber \\  &\to& 
 \frac{A^{N_B}}{q^N} \sum_{\{ \mu_i \}} e^{\frac{\beta_{\rm P} }{q}
 \sum_{\langle i, j \rangle}
 \sum_{r_{i, j} = 1}^{q - 1}  
 J^{( r_{i, j} )}_{i, j} ( \nu_{i, j} ) \mu^{q - r_{i, j}}_i
 \mu^{r_{i, j}}_j } \, .  \label{eq:Pnuij3}
\end{eqnarray}

For the Potts gauge glass model,
 $V (n_b ) $ is given by
\begin{equation}
 V (n_b ) = (e^{\beta J} - 1)^{n_b} e^{- \frac{N_B \beta J}{q}} \, .
 \label{eq:VPGG} 
\end{equation}
The way of deriving Eq.~(\ref{eq:VPGG}) is described in Appendix B.
 For the Potts gauge glass model, $P_{\rm FK}$ is given by \cite{Y2}
\begin{eqnarray}
  P_{\rm FK} (q_i, q_j, \nu_{i, j}) =
 & & 1 - \exp \biggl\{ - \frac{\beta J}{q} \biggl[
 \sum_{r_{i, j} = 1}^{q - 1} e^{\frac{2 \pi i}{q} (\nu_{i, j} + q_i - q_j) r_{i, j}} + 1
 \biggr] \biggr\} \, . \qquad
 \label{eq:PFKPGG}
\end{eqnarray}
 The way of deriving Eq.~(\ref{eq:PFKPGG}) is also described in Appendix B.
 By using the gauge transformation, the $P_{\rm FK}$ part becomes $P_{\rm FK} \to P_{\rm FK}$.

By using Eqs.~(\ref{eq:Z}), (\ref{eq:energy}) and (\ref{eq:VPGG}),
 the internal energy $E$ is given by
\begin{equation}
 E = \frac{N_B  J}{q} - \frac{J e^{\beta J}}{e^{\beta J} - 1}
 [ \langle n_b \rangle_T ]_R \, . \label{eq:energy_PGG}
\end{equation}
 $[\langle n_b \rangle_T ]_R$ is given by 
\begin{equation}
 [ \langle n_b \rangle_T ]_R = \sum_{\langle i, j \rangle} 
 [\langle P_{\rm FK} (q_i, q_j, \nu_{i, j}) \rangle_T ]_R \, . \label{eq:nbTRPGG1}
\end{equation}
 When $\beta = \beta_P$,
 $[\langle P_{\rm FK} (q_i, q_j, \nu_{i, j}) \rangle_T ]_R$ is obtained
 by using the gauge transformation as \cite{Y2}
\begin{eqnarray}
  & & [\langle P_{\rm FK} (q_i, q_j, \nu_{i, j}) \rangle_T ]_R
  \nonumber \\
  &=& \sum_{ \{ \nu_{l, m} \}}
 \prod_{\langle l, m \rangle} P ( \nu_{l, m})
  \frac{\sum_{\{ q_l \} } 
 P_{\rm FK} ( q_i, q_j, \nu_{i, j})
 \, e^{ - \beta_P {\cal H} (\{ q_l \}, \{ \nu_{l, m} \})}}
 {\sum_{\{ q_l \} } 
 e^{- \beta_P {\cal H} (\{ q_l \}, \{ \nu_{l, m} \})}} \nonumber \\
  &=& \frac{A^{N_B}}{q^N} \sum_{ \{ \nu_{l, m} \}}
 \sum_{\{ q_l \} } P_{\rm FK} (q_i, q_j, \nu_{i, j})
 \, e^{- \beta_P {\cal H} (\{ q_l \}, \{ \nu_{l, m} \})} \nonumber \\ 
 &=& \frac{e^{\beta_P J} - 1}{e^{\beta_P J} + q - 1 } \, ,
  \label{eq:PFKTRPGG}
\end{eqnarray}
 where $\beta_P$ is the inverse temperature on the Nishimori line.
 By using Eqs.~(\ref{eq:nbTRPGG1}) and (\ref{eq:PFKTRPGG}),
 we obtain
\begin{equation}
 [ \langle n_b \rangle_T ]_R
 = \frac{N_B (e^{\beta_P J} - 1)}{e^{\beta_P J} + q - 1 } \, .
 \label{eq:nbTRPGG2}
\end{equation}
 By using Eqs.~(\ref{eq:energy_PGG}) and (\ref{eq:nbTRPGG2}), 
 the internal energy $E$ is obtained as
\begin{equation}
 E = \frac{N_B  J}{q} - \frac{N_B J e^{\beta_P J}}{e^{\beta_P J} + q - 1} \, .
\end{equation}
 This solution is exact, and is equivalent to
 the solution in Ref.~\citen{NS}.

By using Eqs.~(\ref{eq:Z}), (\ref{eq:C}) and (\ref{eq:VPGG}),
 the specific heat $C$ is given by
\begin{eqnarray}
 C &=& k_B \biggl( \frac{\beta J}{2} \biggr)^2
  {\rm cosech}^2 \biggl( \frac{\beta J}{2} \biggr)
 \{ - [\langle n_b \rangle_T ]_R \nonumber \\
 &+& e^{\beta J} ([\langle n_b^2 \rangle_T ]_R
 - [ \langle n_b \rangle_T^2 ]_R ) \} \, .
 \label{eq:CPGG}
\end{eqnarray}
 $[\langle n_b^2 \rangle_T ]_R$ is given by
\begin{eqnarray}
 & & [\langle n_b^2 \rangle_T ]_R \nonumber \\
 &=& \sum_{\langle i, j \rangle} \sum_{\langle k, l \rangle}
 [\langle P_{\rm FK} (q_i, q_j, \nu_{i, j})
 P_{\rm FK} (q_k, q_l, \nu_{k, l})
 (1 - \delta_{i, k} \delta_{j, l} - \delta_{i, l} \delta_{j, k}  )
 \nonumber \\
 &+& P_{\rm FK} (q_i, q_j, \nu_{i, j})
 (\delta_{i, k} \delta_{j, l} + \delta_{i, l} \delta_{j, k} ) \rangle_T ]_R \, .
\end{eqnarray}
 By performing a similar calculation with 
 the calculation in Eq.~(\ref{eq:PFKTRPGG}), we obtain
\begin{eqnarray}
 [\langle n_b^2 \rangle_T ]_R &=& \frac{N_B (N_B - 1) (e^{\beta_P J} - 1)^2}
{(e^{\beta_P J} + q - 1)^2} \nonumber \\
 &+& \frac{N_B (e^{\beta_P J} - 1)}
{e^{\beta_P J} + q - 1} \, . \label{eq:nb2TRPGG}
\end{eqnarray}
 By applying the Cauchy-Schwarz inequality, we obtain
\begin{equation}
 [ \langle n_b \rangle^2 ]_R \ge [ \langle n_b \rangle ]^2_R =
 \frac{N^2_B (e^{\beta_P J} - 1)^2}{(e^{\beta_P J} + q - 1)^2} \, .
 \label{eq:nb2TR2PGG}
\end{equation}
 Therefore, by using Eqs.~(\ref{eq:nbTRPGG2}),
 (\ref{eq:CPGG}), (\ref{eq:nb2TRPGG}) and (\ref{eq:nb2TR2PGG}),
 we obtain the upper bound of the specific heat $C$ as
\begin{equation}
 C \le \frac{k_B N_B (\beta_P J )^2 e^{\beta_P J} (q - 1)}
 {(e^{\beta_P J } + q - 1)^2} \, .
\end{equation}
 This solution is rigorous, and is equivalent to
 the solution in Ref.~\citen{NS}.

\section{Summary} \label{sec:5}

We derived the 
 exact internal energies and the rigorous upper bounds of 
 specific heats for the $\pm J$ Ising model and 
 a Potts gauge glass model by applying
 the gauge theory to the Fortuin-Kasteleyn representation.
 The results were derived on the Nishimori lines.
 The present solutions agreed with the previous solutions
 in Refs.~\cite{N, MH, NS}.
 The Fortuin-Kasteleyn
 representation is a representation based on a percolation picture for
 spin-spin correlation.
 The derivation of the solutions by the present method
 must be useful for understanding the relationship
 between the percolation picture for spin-spin correlation 
 and the physical quantities on the Nishimori line.

\appendix
 
\section{The weight and the probability for active bond in the $\pm J$ Ising model}

We will derive Eqs.~(\ref{eq:VpmJ}) and (\ref{eq:PFKpmJ}).
 The framework for 
 the way to derive Eqs.~(\ref{eq:VpmJ}) and (\ref{eq:PFKpmJ}) is
 described in Ref.~\citen{KG}.
 We define the weight of two spins as $w (S_i, S_j, J_{i, j})$. 
 $w (S_i, S_j, J_{i, j})$ is given by
\begin{equation}
 w (S_i, S_j, J_{i, j}) 
 = \exp ( \beta J_{i, j} S_i S_j ) \, .
 \label{eq:aa-1}
\end{equation}
We define the weight for $J_{i, j} S_i S_j = J$
 as $w_{\rm para}$. We obtain
\begin{equation}
 w_{\rm para} (S_i, S_j, J_{i, j}) 
 = \exp ( \beta J )  \, . \label{eq:aa-2}
\end{equation}
We define the weight for $J_{i, j} S_i S_j = - J$
 as $w_{\rm anti}$. We obtain
\begin{equation}
 w_{\rm anti} (S_i, S_j, J_{i, j})
  = \exp ( - \beta J )  \, . \label{eq:aa-3}
\end{equation}
 We define the weight of graph for connecting two spins as $w (g_{\rm conn})$.
 We define the weight of graph for disconnecting two spins
 as $w (g_{\rm disc})$.
 We are able to write 
\begin{eqnarray}
 w_{\rm para} (S_i, S_j, J_{i, j})
  &=& w (g_{\rm conn}) + w (g_{\rm disc}) \, , \label{eq:aa-5} \\
 w_{\rm anti} (S_i, S_j, J_{i, j})
 &=& w (g_{\rm disc})  \, . \label{eq:aa-6} 
\end{eqnarray}
 By using Eqs.~(\ref{eq:aa-2}), (\ref{eq:aa-3}), (\ref{eq:aa-5}) and (\ref{eq:aa-6}),
 we obtain
\begin{eqnarray}
 w (g_{\rm conn}) 
 &=& \exp ( \beta J ) -  \exp ( - \beta J ) \, , \label{eq:aa-7} \\
 w (g_{\rm disc})
 &=&  \exp ( - \beta J )  \, . \label{eq:aa-8}
\end{eqnarray}
 By using Eqs.~(\ref{eq:aa-7}) and (\ref{eq:aa-8}),
 we obtain
 the weight $V (n_b )$ for the active bond number $n_b$ as
\begin{equation}
 V (n_b ) = ( e^{\beta J} - e^{- \beta J } )^{n_b } \, 
 ( e^{- \beta J } )^{N_B - n_b}
\end{equation}
 The above equation is equal to Eq.~(\ref{eq:VpmJ}).
 We define the probability of
 connecting two spins for $J_{i, j} S_i S_j = J$ as
 $P_{\rm para} (g_{\rm conn})$.
 We define the probability of
 connecting two spins for $J_{i, j} S_i S_j = - J$
 as $P_{\rm anti} (g_{\rm conn})$.
 We are able to write
\begin{eqnarray} 
 P_{\rm para} (g_{\rm conn})
 &=& \frac{w (g_{\rm conn})}{w (g_{\rm conn}) + w (g_{\rm disc})} \,
 , \label{eq:aa-9} \\
 P_{\rm anti} (g_{\rm conn})
 &=& 0 \, . \label{eq:aa-10}
\end{eqnarray} 
 By using Eqs.~(\ref{eq:aa-7}), (\ref{eq:aa-8}), (\ref{eq:aa-9}), (\ref{eq:aa-10}), 
 we derive Eq.~(\ref{eq:PFKpmJ}).

\section{The weight and the probability for active bond in a Potts gauge glass model}

We will derive Eqs.~(\ref{eq:VPGG}) and (\ref{eq:PFKPGG}).
 The framework for 
 the way to derive Eqs.~(\ref{eq:VPGG}) and (\ref{eq:PFKPGG}) is
 described in Ref.~\citen{KG}.
 We define the weight of two spins as $w (q_i, q_j, \nu_{i, j})$. 
 $w (q_i, q_j, \nu_{i, j})$ is given by
\begin{eqnarray}
 & & w (q_i, q_j, \nu_{i, j}) \nonumber \\
 &=& \exp \biggl\{ \frac{\beta J}{q}
 \sum_{r_{i, j} = 1}^{q - 1}
 \exp \biggl[ \frac{2 \pi i}{q}
 \biggr( \nu_{i, j} + q_i - q_j \biggl) r_{i, j}
 \biggr] \biggr\} \, .  \label{eq:ab-1}
\end{eqnarray}
We define the weight for $\nu_{i, j} + q_i - q_j = 0$
 as $w_{\rm para}$. We obtain
\begin{equation}
 w_{\rm para} (q_i, q_j, \nu_{i, j}) 
 = \exp \biggl[ \frac{\beta J (q - 1) }{q} \biggr] \, . \label{eq:ab-2}
\end{equation}
We define the weight for $\nu_{i, j} + q_i - q_j \ne 0$
 as $w_{\rm anti}$. We obtain
\begin{equation}
 w_{\rm anti} (q_i, q_j, \nu_{i, j})
  = \exp \biggl( - \frac{\beta J}{q} \biggr) \, . \label{eq:ab-3}
\end{equation}
 We define the weight of graph for connecting two spins as $w (g_{\rm conn})$.
 We define the weight of graph for disconnecting two spins
 as $w (g_{\rm disc})$.
 We are able to write 
\begin{eqnarray}
 w_{\rm para} (q_i, q_j, \nu_{i, j})
  &=& w (g_{\rm conn}) + w (g_{\rm disc}) \, , \label{eq:ab-5} \\
 w_{\rm anti} (q_i, q_j, \nu_{i, j})
 &=& w (g_{\rm disc})  \, . \label{eq:ab-6} 
\end{eqnarray}
 By using Eqs.~(\ref{eq:ab-2}), (\ref{eq:ab-3}), (\ref{eq:ab-5}) and (\ref{eq:ab-6}),
 we obtain
\begin{eqnarray}
 w (g_{\rm conn}) 
 &=& \exp \biggl[ \frac{\beta J (q - 1) }{q} \biggr] 
 -  \exp \biggl( - \frac{\beta J}{q} \biggr) \, , \label{eq:ab-7} \\
 w (g_{\rm disc})
 &=&  \exp \biggl( - \frac{\beta J}{q} \biggr)  \, . \label{eq:ab-8}
\end{eqnarray}
 By using Eqs.~(\ref{eq:ab-7}) and (\ref{eq:ab-8}),
 we obtain
 the weight $V (n_b )$ for the active bond number $n_b$ as
\begin{equation}
 V (n_b ) = [ e^{ \frac{\beta J (q - 1) }{q} } 
 -  e^{ - \frac{\beta J}{q} } ]^{n_b } \, 
 ( e^{ - \frac{\beta J}{q} } )^{N_B - n_b} \, .
\end{equation}
 The above equation is equal to Eq.~(\ref{eq:VPGG}).
 We define the probability of
 connecting two spins for $\nu_{i, j} + q_i - q_j = 0$ as
 $P_{\rm para} (g_{\rm conn})$.
 We define the probability of
 connecting two spins for $\nu_{i, j} + q_i - q_j \ne 0$
 as $P_{\rm anti} (g_{\rm conn})$.
 We are able to write
\begin{eqnarray} 
 P_{\rm para} (g_{\rm conn})
 &=& \frac{w (g_{\rm conn})}{w (g_{\rm conn}) + w (g_{\rm disc})} \,
 , \label{eq:ab-9} \\
 P_{\rm anti} (g_{\rm conn})
 &=& 0 \, . \label{eq:ab-10}
\end{eqnarray}
 By using Eqs.~(\ref{eq:ab-7}), (\ref{eq:ab-8}), (\ref{eq:ab-9}) and (\ref{eq:ab-10}), 
 we derive Eq.~(\ref{eq:PFKPGG}).

\end{document}